\documentclass[aps,prstb,preprint,onecolumn,showpacs,groupedaddress]{revtex4-1}

\usepackage{graphicx}
\usepackage{epstopdf}
\usepackage{amsmath}
\usepackage{url}
\usepackage{bm}

\begin{document}
\title{Terahertz radiation source using an industrial electron linear accelerator}

\author{Yashvir Kalkal}
\email{yashvirkalkal@gmail.com}
\author{Vinit Kumar}
\email{vinit@rrcat.gov.in}

\affiliation{Homi Bhabha National Institute, Mumbai 400094, India \\
Accelerator and Beam Physics Laboratory, Raja Ramanna Centre for Advanced Technology, Indore 452013, India}

\begin{abstract}
High power ($\sim 100$ kW) industrial electron linear accelerators (linacs) are used for irradiation applications e.g., for pasteurization of food products, disinfection of medical waste, etc. We propose that high power electron beam from such an industrial linac can be first passed through an undulator to generate powerful terahertz (THz) radiation, and the spent electron beam coming out of the undulator can still be used for industrial applications. This will enhance the utilisation of a high power industrial linac. We have performed calculation of spontaneous emission in the undulator to show that for typical parameters, continuous terahertz radiation having power of the order of $\mu$W can be produced, which may be useful for many scientific applications.
\end{abstract}


\maketitle

\section{Introduction}\label{sec:1}
Due to several important scientific applications in research and in industry, there is a demand for powerful and tunable sources of terahertz (THz) radiation~\cite{Seigel,THz,THzimaging}. Direct laser based sources e.g., optically pumped lasers and quantum cascade lasers provide THz radiation with average output power of the order of tens or hundreds of mW~\cite{THzMueller}. These sources are however inherently not continuously tunable. Conventional sources such as parametric oscillators and time-domain systems give pulsed THz radiation with relatively low average output power ($\sim$tens of nW)~\cite{THzMueller}. On the other hand, electron beam based sources e.g., backward wave oscillators (BWOs) and free-electron lasers (FELs) are very promising and powerful sources of continuously tunable THz radiation. Commercially available BWOs produce milliwatts of output power with maximum operating frequency upto 1 THz~\cite{BWOPRL}. FELs, where an electron beam is passed through an undulator immersed in a resonating cavity for the generation of coherent radiation, are known for their wide tunability~\cite{science}. THz radiation is produced in an FEL due to interaction of the electron beam with the on-axis, static transverse magnetic field, varying sinusoidally along the undulator axis, in the presence of electromagnetic field building up in the resonator cavity. Lasing in these devices however requires high peak current; which needs an electron beam of very short pulse duration ($\sim$ picoseconds)~\cite{FELps}. A high quality electron beam with very low energy spread is required to achieve lasing in an FEL system~\cite{Braubook}. The infrastructure needed to meet these requirements is quite expensive and bulky; which makes these devices  impractical to use for the table top experiments. Investigations on the compact electron beam based sources of THz radiation, such as Smith-Purcell FELs~\cite{VinitPRE,SPFEL1} and $\check{\text{C}}$erenkov FELs~\cite{Walsh3,CFEL} are attractive and promising. These devices however require very low emittance electron beam of low energy, which are yet to be demonstrated~\cite{CFEL,KimPRSTB}.

Recently, there has been lot of interest in making high average power (up to 100 kW) industrial electron linear accelerator (linac) for various industrial applications such as radiotherapy, polymer reforming and materials irradiation~\cite{BARC,LINAC30,ILU1}. A high energy ($\sim$ 10 MeV) electron beam is favourable for the irradiation processes due to its high penetration depth. Note that a low average power electron beam with very fine energy spread and emittance generates powerful THz radiation in an FEL system, due to coherent stimulated emission. The quality of the electron beam from a typical industrial linac may not be very good for the operation of a free-electron laser, but when such a high average power electron beam passes through an undulator, it can emit powerful THz radiation through spontaneous emission. This radiation can fulfil the requirements of many scientific applications, such as imaging of biological samples, inspecting packaging and analysing chemical composition~\cite{THzimaging,THzTDS, THzimagingCW}. Thereafter, the spent electron beam can be used for the irradiation application, which gives us twofold advantage. An experimental observation of THz radiation from undulator through spontaneous emission by using a 2 mA, 7.5 MeV electron beam has been reported recently in Ref.~\cite{THzexp}. In this paper, we have performed calculations to estimate the power of emitted spontaneous radiation, when an electron beam emerging from a high average power industrial linac dedicated for the irradiation application, is passed through an undulator with optimized parameters.

\begin{figure}[t]
\includegraphics[width=16.5 cm]{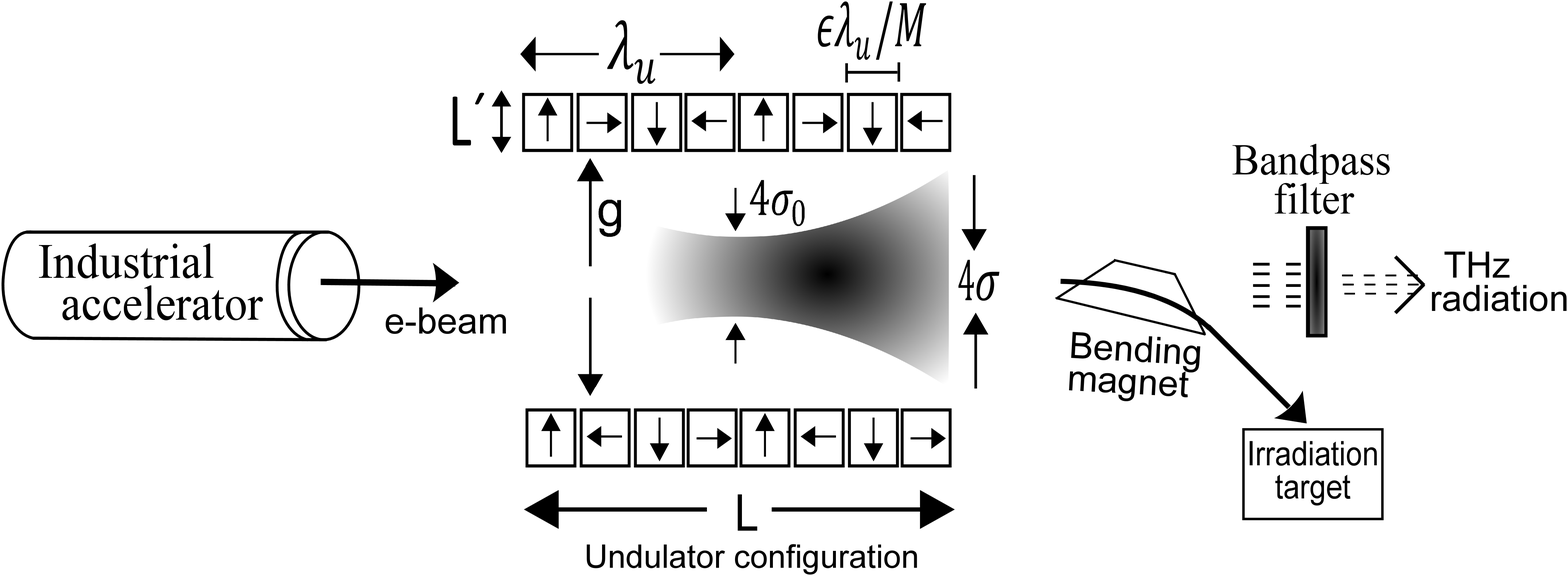}
\caption{Schematic of a device based on the arrangement of industrial linac and undulator, to produce THz radiation along with the irradiation application.}
\label{fig:fig1}
\end{figure}
\section{Theoretical analysis}\label{sec:2}
A schematic of the device is shown in Fig.~\ref{fig:fig1}. A high energy electron beam emerging from a powerful industrial linac is allowed to pass through an undulator, before being directed on the target to be irradiated. It will generate copious THz radiation in the undulator through spontaneous emission. Here, we will perform an analysis for the calculation of power radiated in the spontaneous emission. The undulator is assumed to consist of $N_u$ number of periods, having a period length $\lambda_u$. The detailed analysis for the motion of electron beam in the undulator is already available in the literature~\cite{Braubook,Kim1,Formfactor,Jackson,Atwood,Kimundulator}, and we can write the expression for the energy radiated per unit frequency width d$\omega$ per unit solid angle d$\Omega$ by the system of $N$ electrons in a bunch, along the direction of unit vector $\boldsymbol{n}$ as~\cite{Formfactor}: 
\begin{eqnarray}
\label{eq:energy}
\frac{d^2\mathcal{I}}{d\omega d\Omega}&=&[N+(N^2-N)f(\omega)]\frac{e^2\omega^2}{16\pi^3\epsilon_0c^3}\Bigg\vert\int\limits_{-\infty}^{\infty}\boldsymbol{n}\times(\boldsymbol{n}\times \boldsymbol{v}) e^{i(\omega t-k_R\boldsymbol{n}.\boldsymbol{r})}dt\Bigg\vert^2.
\end{eqnarray}
Here, $\omega=ck_R$ is the angular frequency of the emitted radiation, $k_R$ is the wavenumber of light, $c$ is the speed of light, $e$ is the electronic charge, $\epsilon_0$ is permittivity of the free space, $\boldsymbol{r}$ is the location of the electron bunch centre, $\boldsymbol{v}$ represents the instantaneous velocity of the particle bunch, $f(\omega)$=$\vert \int e^{i\omega r/c}S(\boldsymbol{r})d^3r\vert^2$ is a form factor, which describes coherence of the emitted light, and $S(\boldsymbol{r})$ is a continuous normalized density distribution function of the electron bunch such that the factor $NS(\boldsymbol{r})d^3r$ gives probability of finding an electron in the region $d^3r$ around $\boldsymbol{r}$. The total power radiated by an electron beam passing through an undulator will depend upon the bunch length of the electron beam. If the electron bunch length is shorter than the wavelength of the light, then the form factor $f(\omega)$=1 represents the coherent limit and the emitted power will be $N^2$ times the result from a single electron~\cite{Formfactor}. For the bunch length of the electron beam significantly greater than the wavelength of the light; form factor $f(\omega)$=0 represents the incoherent limit and the power radiated by the electron beam will be $N$ times the result of single electron~\cite{Formfactor}. Hence, a pre-bunched electron beam with bunch length smaller than the wavelength of light can generate high power, coherent THz radiation, when passed through an undulator\cite{THzE,THz1}. This scheme has been successfully used in a Compact Advanced Terahertz Source (CATS) at ENEA, Italy to generate THz radiation~\cite{Enea}. However, this approach requires electron beam with low energy spread, and an additional RF cavity to produce a pre-bunched electron beam. Moreover, the performance is critically dependent on the shape of the electron bunch~\cite{Prebs,Preb}. 

The finite energy spread of the electron beam also affects the radiation emitted when the electron beam passes through an undulator. In the case of lasing in an FEL, the gain profile of stimulated emitted radiation is very sensitive to electron energy and therefore, one requires a high quality electron beam with low energy spread to maintain the resonance condition of the system~\cite{Energyspread1,Energyspread2}. In the case of spontaneous emission of radiation from the undulator, if the energy spread of the electron beam is large, the intensity spectrum will be broad; keeping the total output power nearly the same~\cite{mona}.  The relative frequency width $\delta \omega/\omega$ due to the energy spread is $2\delta\gamma/\gamma$, where $\gamma$ is the energy of the electron beam in unit of its rest mass energy. The total power radiated in all harmonics in an undulator, integrated over all wavelengths, and angles is given by $P_T=\pi e\gamma^2 IN_uK^2/3\epsilon_0\lambda_u$~\cite{Kim1,Atwood}. Here, $I$ is the beam current, $K$=$eB_0/k_umc$ is the peak value of the undulator parameter, $B_0$ represents peak undulator magnetic field, $k_u$=$2\pi/\lambda_u$, and $m$ is mass of the electron. For the spectroscopy and imaging related applications, we want narrow spectral bandwidth of the radiation. In the undulator, narrower bandwidth radiation is emitted inside the central radiation cone, with semi-angle $\theta=1/\gamma\sqrt{N_u}$, around the beam axis~\cite{Atwood}; which can be selected by inserting a bandpass filter in the passage of the output radiation as shown in Fig.~\ref{fig:fig1}. In the central radiation cone, the average power radiated at the fundamental frequency due to the spontaneous emission by $N$ electrons for an arbitrary $K$ is given by~\cite{Kim1,Atwood}:
\begin{eqnarray}
\label{eq:power}
P_{cen}&=&\frac{\pi e \gamma^2I}{\epsilon_0\lambda_u}\frac{f(K)K^2}{(1+K^2/2)^2},
\end{eqnarray}
where $f(K)=[J_0(x)-J_1(x)]^2$, and $x=K^2/{(4+2K^2)}$.

Another important quantity of the radiation in the imaging related applications is spectral brightness, which is defined as the photon flux per unit area and per unit solid angle at the source within a relative bandwidth of 0.1 $\%$. The expression for the on-axis spectral brightness of the undulator radiation is given by~\cite{Kim1,Atwood}
\begin{equation}
\label{eq:brightness}
\hspace*{-15pt}\mathcal{B}_{(\Delta \omega/\omega)}=\frac{7.25\times 10^6\times\gamma^2I(\text{A})N_u^2K^2f(K)}{\sigma_e^2(\text{mm}^2)\big(1+\sigma_e^{'2}/\theta^2\big)\big(1+K^2/2\big)^2}\frac{\text{photons/s}}{\text{mm$^2$mrad$^2$ (0.1 $\%$ BW)}},
\end{equation}
where $I$ is in amperes, $\sigma_e$ is the electron beam size in millimeters, and $\sigma_e'$ is the rms beam divergence. Note that above formula is valid for the monochromatic electron beam having very small divergence i.e., $\sigma_e^{'}\ll\theta$. For the case $\sigma_e^{'}\simeq\theta$, Eq.~(\ref{eq:brightness}) overestimates the brightness by a factor of 2, and we need to use numerical simulations for more accurate estimate of the spectral brightness~\cite{Atwood}.

The emitted radiation from the undulator will propagate along the direction of the electron beam. The radiation beam can be out-coupled by putting a window at the end of the interaction region, and can be used in various experiments. The radiation beam will undergo diffraction in the transverse direction and one has to make sure that the diffracting optical beam should not be distorted by the vacuum chamber of the undulator. A rigorous analysis for the representation of the undulator radiation has been recently given by Lindberg and Kim~\cite{Kimundulator}, where the radiation beam coming out of the undulator is described as a freely diffracting beam having rms beam waist size $\sigma_{0} = \sqrt{\lambda_R L} /2\pi$, and rms beam divergence as $\sigma_{o}'$=(1/2)$\sqrt{\lambda_R/L}$. Here, $L$ is the length of the undulator, and $\lambda_R = \lambda_u(1+K^2/2)/2\gamma^2$ is the on axis central wavelength of the radiated spectrum. It can be assumed that the beam waist is formed at the centre of the undulator. As we move away from the centre of the undulator, the beam size increases due to diffraction, which is described here in terms of Rayleigh range $Z_R = L/\pi$~\cite{Kimundulator}. The rms beam size $\sigma$ at the exit of the undulator is given by $\sigma = \sigma_0 \sqrt{1+(L/2Z_R)^2}$. Note that this calculation assumes an electron beam of negligible size and divergence. Taking the finite size and divergence of the electron beam, the formulae for the rms beam size at the waist and the rms beam divergence get modified as $\sigma_0 = \sqrt{\lambda_R L /4\pi^2 + \sigma_e^2}$, and $\sigma_{o}'$=$\sqrt{\lambda_R/4L+\sigma_e'^2}$ respectively~\cite{Kim1}. We would like to mention that the effect of finite electron beam size and divergence is not very significant if the unnormalised rms electron beam emittance $\varepsilon_{un}$, which can be understood as product of rms electron beam waist size and divergence, is much less than $\lambda_R/4\pi$, which is the product of rms optical beam waist size and divergence~\cite{Kim1}. In terms of normalized rms beam emittance $\varepsilon_n = \beta \gamma \varepsilon_{un}$, this criterion can be expressed as $\varepsilon_n \ll \beta \gamma \lambda_R/4\pi$, where $\beta$ is the speed of electron in unit of speed of light.

The undulator gap $g$ has to be chosen sufficiently greater than the total beam diameter at the undulator exit, i.e., four times the rms beam size $\sigma$. Here, one has to keep in the mind that increase in $g$ value will diminish the on-axis undulator magnetic field since it is  proportional to $\exp{(-k_ug)}$~\cite{Braubook}. One needs to optimize the magnetic field strength together with the undulator length such that the beam diameter at the undulator exit remains less than the undulator gap $g$. The expression for the peak undulator field  $B_0$ for the Halbach configuration of a pure permanent magnet (PPM) based undulator is given by~\cite{Braubook}:
\begin{eqnarray}
\label{eq:field}
B_0 &=& 2B_{rem}e^{-k_ug/2}(1-e^{-k_uL'})\frac{\sin{(\epsilon\pi/M)}}{\pi/M}.
\end{eqnarray}
Here, $B_{rem}$ is the remnant field of the PPM, and has value 1.1 T for NdFeB magnets; $M$ is the number of magnets required to complete one period and $L'$ represents the transverse length of the magnet. In the most common configuration, $M$=4, $\epsilon$=1 and $L'$=$\lambda_u/4$. Using these values in Eq.~(\ref{eq:field}), we obtain the peak magnetic field in an undulator as~\cite{Braubook}:
\begin{eqnarray}
B_0&=&1.57\times\exp{\big(-\frac{\pi g}{\lambda_u}\big)}.
\end{eqnarray} 
The expression for the peak value of undulator parameter $K$ is given by:
\begin{eqnarray}
\label{eq:k}
K &=& 1.48\times\lambda_u(cm)\times \exp{\big(-\frac{\pi g}{\lambda_u}\big)}.
\end{eqnarray}
The analysis which we have performed in this section for the spontaneous emission of THz radiation in an undulator will be helpful in obtaining the parameters of a practical device, as described in the following section.

\section{An example case}
To perform the calculations, we now take an example case of a high average power industrial linac, and optimize the parameters of undulator in accordance with the analysis described in Sec.~\ref{sec:2}. We consider electron beam linac having electron beam energy in the range 7.5 to 10 MeV, average beam current of 10 mA, and average power upto 100 kW. The relative energy spread $\delta\gamma/\gamma$ is assumed to be around 7 $\%$. The chosen parameters are close to the parameters of a S-band linac discussed in Ref.~\cite{ILU1}.

For a given energy spread of the electron beam from linac, we choose number of undulator periods $N_u\simeq \gamma/\delta\gamma$. Based upon this argument, we have chosen~$N_u$= 15 in our calculations. Note that if we take the undulator length more than this value, the spectrum width, and the brightness of emitted radiation will be limited by the energy spread of the electron beam. Also, if we take a longer undulator, the radiation beam size at the exit of the undulator will increase and the radiation beam may strike the edge of the undulator. We have taken undulator period $\lambda_u$ as 50 mm and undulator parameter $K$ from 0.6 to 2.1. These values of $K$ parameter can be achieved by choosing the undulator gap $g$ between 40 to 20 mm of an undulator made up of pure permanent magnets~\cite{Braubook}.

\begin{figure}[t]
\begin{center}
\includegraphics[width=16.5 cm]{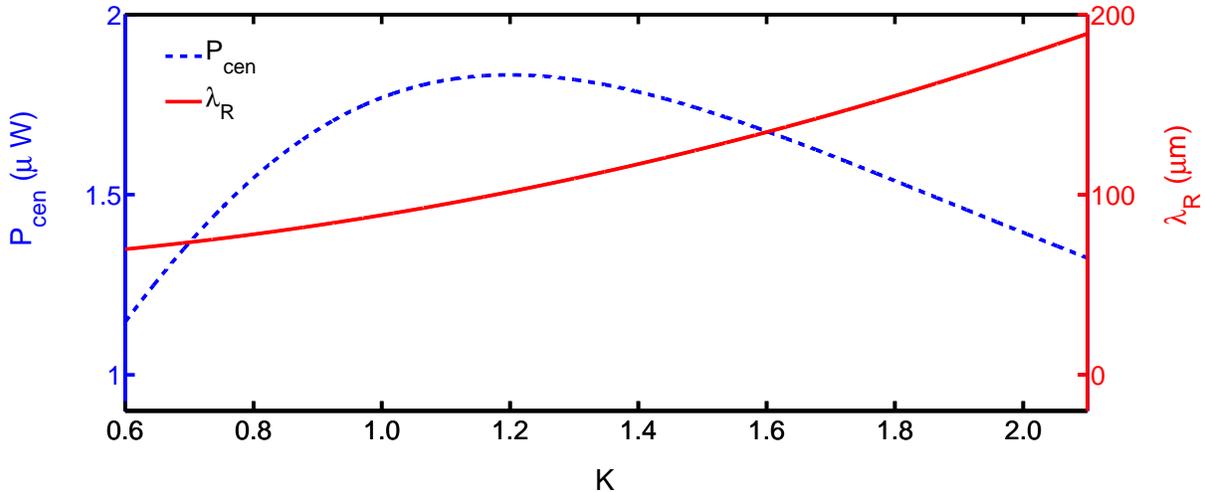}
\caption{Plot of output power (dashed) in central cone and operating wavelength (solid) as a function of undulator parameter $K$ for $E$ = 10 MeV, $I$ = 10 mA, $N_u$ = 15, and $\lambda_u$=50 mm.}
\label{fig:fig2}
\end{center}
\end{figure} 

In Fig.~\ref{fig:fig2}, we have shown the variation of the output power in the central cone and the radiation wavelength as a function of undulator parameter $K$ for a 10 MeV electron beam. The parameters used in the calculations have been listed in Table 1. The radiation wavelength increases with $K$ value, and the output power shows a maxima near $K$=1.2. The maximum output power in the central radiation cone is obtained as 1.8 $\mu$W at 3 THz frequency. The selection of narrow spectrum around the central wavelength can be made by using a THz bandpass filter in the passage of output radiation. The bandpass filters fabricated from gold-mesh frequency-selective surfaces are commercially available, and have transmission of about 80 $\%$ at 3 THz frequency~\cite{Bandpass}. The filtered radiation will have a continuous average power of 1.5 $\mu$W, and can be transported into a nearby experimental station via suitable optical arrangements. We would like to mention that average power of around tens of nW can be achieved in conventional THz sources such as parametric oscillators, photo-mixing and time domain systems~\cite{THzMueller}, however, all these sources are not continuously tunable. Further, the relative bandwidth of the output radiation at central wavelength in our system is around 14 $\%$, which is nearly three times less than the relative bandwidth of output radiation in the conventional sources described above.
\begin{table}[b]
\caption{Parameters of the undulator}
\label{tab:table2}
\smallskip
\centering
\begin{tabular}{|lc|}
\hline
Undulator period ($\lambda_u$) &  50 mm\\
No. of periods ($N_u$) & 15\\
Undulator gap ($g$) & 20 - 40 mm\\
Undulator parameter ($K$) &2.1 - 0.6 \\
Peak magnetic field ($B_0$) & 0.45 - 0.13 T\\
Radiation wavelength ($\lambda_R$) & 190 - 70 $\mu m$ \\
\hline
\end{tabular}
\end{table}

\begin{figure}[t]
\centering
\includegraphics[width=16.5 cm]{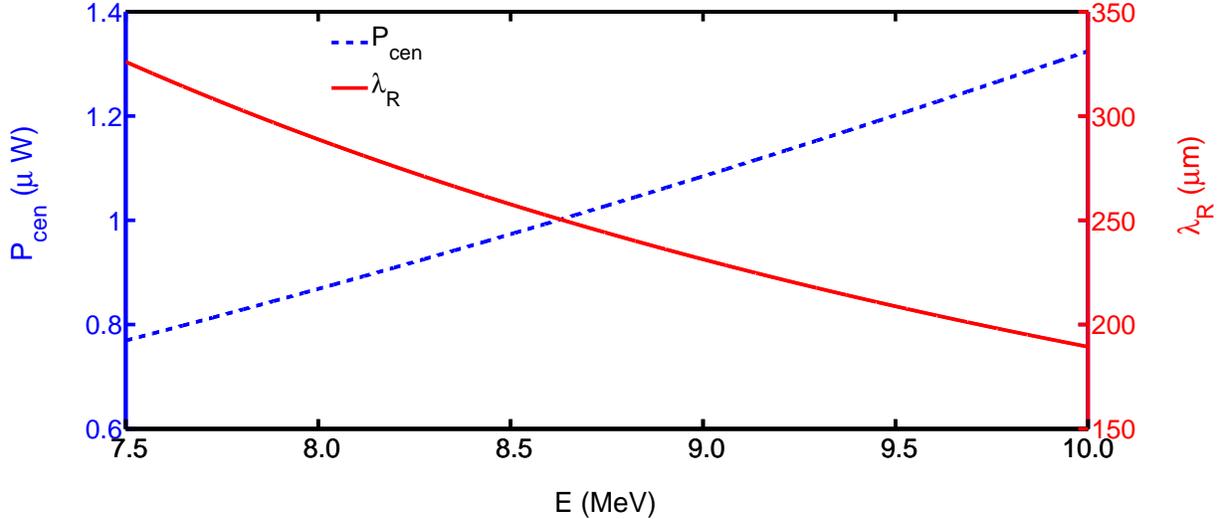}
\caption{Plot of output power (dashed) in central cone and operating wavelength (solid) as a function of electron beam energy for $I$ = 10 mA, $K$ = 2.1, $\lambda_u$ = 50 mm, and $N_u$ = 15.}
\label{fig:fig3}
\end{figure}

For a fixed undulator period, the wavelength of the output radiation in the proposed system can be tuned by either changing the electron beam energy or by changing the undulator gap. In our example case, we take energy range from 7.5 MeV to 10 MeV, and the undulator gap $g$ can be changed from 20 mm to 40 mm under normal tuning range. As shown in Fig. 2, radiation wavelength can be tuned from 70 $\mu$m to 190 $\mu$m by varying $K$ from 0.6 to 2.1 for a 10 MeV electron beam. For $E=10$ MeV and $K=2.1$, the rms radiation beam waist size $\sigma_0$ is calculated as 2 mm by taking the account of finite electron beam size as mentioned in the previous section. The radiation beam diameter at the exit of undulator is obtained as 15.3 mm. In these calculations, we have taken the normalized electron beam emittance as 30 mm-mrad, which can be easily achieved in a typical 10 MeV electron linac~\cite{LINAC}. Note that the effect of finite electron beam size and divergence will not be significant as long as the normalized rms electron beam emittance is less than $\beta \gamma \lambda_R/4\pi$, which is 310 mm-mrad for $\lambda_R=190~\mu$m. We have also found that the space charge effects are negligible for the chosen parameters of the electron beam as the emittance term turns out to be much smaller then the space-charge term in the beam envelope equation~\cite{Chao}. This condition can be expressed as~\cite{Chao}:
\begin{eqnarray}
\label{eq:k}
\hspace*{100pt}\frac{\sigma_e^2}{2\beta\gamma\varepsilon_n^2}\frac{I_p}{I_A}&<&1,
\end{eqnarray}
where $I_p=4.8$ A is the peak current of micropulse, and $I_A=17.04$ kA is the Alfv$\acute{\text{e}}$n current. We evaluate left-hand side of above equation, and obtain 0.003; for which the inequality is satisfied. The undulator gap $g$ has to be larger than the optical beam size and we have $g$=20 mm. For $K$=2.1, the radiation wavelength can be tuned further towards higher values by decreasing the electron beam energy, as shown in Fig.~\ref{fig:fig3}. Further increment in $K$ value will make the optical beam size comparable or larger than the undulator gap. By using these parameters, we can get output radiation ranging from 0.9 THz to 4.3 THz.

An approximate value of spectral brightness $\mathcal{B}_{\Delta\omega/\omega}$ of the emitted radiation for $K$=1.2 and $E$=10 MeV is estimated by using Eq.~(\ref{eq:brightness}) as $9\times 10^8$ photons/s/mm$^2$/ mrad$^2$ (0.1$\%$ BW), which can also be written as $\mathcal{B}_{\Delta\omega/\omega}=0.6\times 10^{-6}$ W/mm$^2$/sr (0.1$\%$ of BW). The spectral brightness of the output radiation in the proposed system is comparable to the brightness of coherent synchrotron radiation at the meterology light source~\cite{THzbrightness}, and higher than the conventional thermal sources. With the help of continuously tunable, and high brightness THz wave, one can perform multispectral THz imaging to identify the composition of components in the chemical samples~\cite{THzimagingCW}. 

As the electron beam propagates down the undulator, an additional energy spread will be induced due to the quantum fluctuations of the spontaneous undulator radiation~\cite{Inducedspread}. For parameters considered in our example case, the relative energy spread induced by the undulator comes out to be 0.015~$\%$, which is quite low as compared to the initial relative energy spread 7.0~$\%$. Hence, the spent electron beam after emitting THz radiation will still be of good quality, and can be used for the irradiation application.

\section{Discussions and Conclusions}
In this paper, we have discussed a proposal to enhance the utilization of high average power industrial electron linac by using an undulator to generate terahertz radiation, in addition to irradiation application. In an FEL, the average power of the electron beam is typically low, which is usually of the order of few tens or hundreds of Watts. Such an electron beam typically generates THz radiation of nW power through spontaneous emission, which gets enhanced to the order of few tens or hundreds of mW through the process of stimulated emission. To have a strong interaction between the electron beam and the optical beam in an FEL system such that it generates stimulated radiation, the electron beam envelope has to well inside the optical beam envelope throughout the interaction length, and its average electron beam radius over the length of the undulator needs to be minimum. This requires that the parameters related to the profile of the electron beam are suitably matched at the entrance of the undulator. This is achieved by having a suitable beam transport line between the linac and the undulator. In addition, as discussed, very high quality electron beam with low energy spread, low emittance and high peak current is needed for the generation of stimulated emission in an FEL. These requirements are not there, when we use high average power electron beam for the generation of terahertz radiation through the process of spontaneous emission, as discussed in the paper. This makes the proposed device simpler.

We would like to emphasize that the length of the undulator needs to be chosen in an optimum manner based on the energy spread of the electron beam. We choose number of undulator periods $N_u\simeq\gamma/\delta\gamma$. A longer undulator will generate brighter radiation according to Eq.~(\ref{eq:brightness}). However, as discussed, Eq.~(\ref{eq:brightness}) is valid for monoenergetic beam. For a beam with finite energy spread, the brightness will reduce since finite energy spread will lead to additional width in radiation spectrum. If we take $\delta\gamma/\gamma>1/N_u$, the brightness and the spectral width of the emitted radiation will be limited by the energy spread of the electron beam. We therefore choose $N_u\simeq\gamma/\delta\gamma$. Also, the radiation beam size at the undulator exit will be large for the longer undulator, and the radiation beam may strike at the edge of the undulator for this case.

To conclude, we have presented an analysis of a device, which is based on the arrangement of a powerful industrial linac and an undulator; to produce useful THz radiation, along with the pre-existing irradiation applications. For the analysis of undulator radiation, we followed a recent approach given in Ref.~\cite{Kimundulator}. By taking an example case of the high average power industrial linac, we have optimized the parameters of an undulator, which can be used to produce copious THz radiation. We observed that an undulator with moderate parameters such as length 0.75 m, period 50 mm and $K$ from 0.6 to 2.1, can be used with 7.5 to 10 MeV, 100 kW linac to produce a continuous tunable THz radiation (0.9 THz to 4.3 THz) with output power in the central cone of the order of $\mu$W. Thus, the utilization of a high power electron linac can be enhanced by putting a short undulator, which may not be significantly expensive. The device can be simultaneously used for terahertz generation, as well as irradiation applications. The output radiation can be tuned by changing the undulator gap or by changing the electron beam energy. Tunable continuous THz radiation is very useful in the imaging and spectroscopy related applications~\cite{THzTDS,THzimagingCW}. Our analysis can be helpful in the detailed optimization of parameters of a practical device.

\acknowledgments
We thank Shankar Lal for useful discussions. We acknowledge Prof. S. B. Roy for constant encouragement. One of us (Y.K.) gratefully acknowledges Homi Bhabha National Institute, Department of Atomic Energy (India) for financial support.


\end{document}